\documentclass[aps,prd,twocolumn,nofootinbib,superscriptaddress,preprintnumbers]{revtex4-1}
\usepackage{amsmath}
\usepackage{subfig}
\usepackage{graphicx}% Include figure files
\usepackage{array}
\usepackage{multirow}
\usepackage{slashed}
\usepackage{amssymb}

\newcommand{\be}{\begin{equation}}
\newcommand{\ee}{\end{equation}}
\newcommand{\bea}{\begin{eqnarray}}
\newcommand{\eea}{\end{eqnarray}}

\font\FermiSmallfont=cmssq8 scaled 1200

\def\UMDppthead#1#2#3{
\null
\begin{center}\vskip -1.0truein{\hbox to 7.5truein {
\hfill
\vbox to 1in {\vfill \FermiSmallfont
              \hbox{#1}
              \hbox{#2}
              \hbox{#3}
              \vfill}
}}\vskip-0.0truein\end{center}}%FNALppthead

\begin{document}

% Page numbers bottom-center
\pagestyle{plain}

%\UMDppthead{UMD-PP-09-034}{}{}

\title{%\begin{flushright} \texttt{UMD-PP-09-034} \end{flushright}
Lower Limits on the Strengths of Gamma Ray Lines\\ from WIMP Dark
Matter Annihilation}

\author{Kevork N.\ Abazajian}
\affiliation{
Maryland Center for Fundamental Physics, Department of Physics,
University of Maryland, College Park, MD 20742}
\affiliation{Department of Physics \& Astronomy, University of California, Irvine, CA 92697}
\author{Prateek Agrawal}
\author{Zackaria Chacko}
\affiliation{
Maryland Center for Fundamental Physics, Department of Physics,
University of Maryland, College Park, MD 20742}
\author{Can Kilic}
\affiliation{
Theory Group, Department of Physics and Texas Cosmology Center, The University of Texas at Austin, Austin, TX 78712}

%\date{\today}

\begin{abstract}

We study the spectra of gamma ray signals that arise from dark matter
annihilation in the universe. We focus on the large class of theories
where the photon spectrum includes both continuum spectrum of gamma rays
that arise from annihilation into Standard Model states at tree level,
as well as monochromatic gamma rays arising from annihilation directly
into two photons at the one loop level. In this class of theories we
obtain lower bounds on the ratio of the strength of the gamma ray line
relative to the gamma ray continuum as a function of the dark matter
mass and spin.  These limits are obtained from the unitarity relation
between the tree level amplitude of the primary annihilation channel and
the imaginary part of the loop level amplitude for annihilation directly
into photons, with the primary decay products running in the loop. These
results are exact in the limit that dark matter annihilation at tree
level is exclusively to a single Standard Model species, occurs through
the lowest partial wave and respects CP. Away from this limit the bounds
are approximate. Our conclusions agree with known results in the
literature for the cases of the Minimal Supersymmetric Standard Model,
Universal Extra Dimensions and the Littlest Higgs with T-Parity. We use
the Fermi-LAT observations to translate these limits into upper bounds
on the dark matter annihilation cross section into any specific Standard
Model state.

\end{abstract}

%\pacs{}
\maketitle
\preprint{UTTG-21-11}
\preprint{TCC-023-11}

%%%%%%%%%%%%%%%%%%%%%%%%%%%%%%%%%%%%%%%%%%%%%%%%%%%%%%%%%%%%%%%%%%%%%%%%%%%%

\section{Introduction}

Cosmological measurements have now established that about 80\% of the
matter in the universe is composed of non-luminous, non-baryonic dark
matter~\cite{Komatsu:2010fb}. However, the precise nature of the
particles of which dark matter is composed remains a mystery. Weakly
Interacting Massive Particles (WIMPs) - thermal relics with weak scale
masses and weak scale cross sections with visible matter - constitute
one well-motivated class of dark matter candidates. In these theories
the relic abundance of the WIMP is set by its annihilation cross
section into Standard
Model (SM) particles, and turns out to be naturally of the right order
to explain observations.

One approach to dark matter detection involves searching for the
products of WIMP annihilation in the universe, such as photons. Dark
matter is constrained to be neutral under electromagnetism, and
therefore in renormalizable theories WIMPs cannot annihilate directly
into photons at tree level. Nevertheless, a continuum spectrum of photons
arises from decays of the primary annihilation products, and also from
final state radiation off charged final states. The spectrum also
includes monochromatic gamma rays that arise from annihilation directly
into $\gamma \gamma$, $\gamma Z$ and $\gamma H$ final states, usually at
loop level.

The continuum spectrum of photons arising from dark matter annihilation
into any specific final state is to a large extent independent of the
detailed form of the matrix element for the process, depending only on
the SM quantum numbers of the particles in the final state and the dark
matter mass. This allows limits on the observed diffuse gamma ray flux
from regions of high dark matter density to be translated into robust
bounds on the rate of dark matter annihilation into any such state. Over
the last few years the Large Area Telescope (LAT) aboard the Fermi
Gamma-ray Space Telescope has been making precise observations of the
gamma ray sky.  Limits on the rate of dark matter annihilation into
various final states have been placed by Fermi-LAT measurements of the
gamma ray flux from nearby galaxy
clusters~\cite{Ackermann:2010rg,*Pinzke:2011ek}, dwarf galaxies
\cite{Scott:2009jn,*Abdo:2010ex,GeringerSameth:2011iw,*Ackermann:2011wa},
the Galactic center~\cite{Hooper:2011ti} and
subhalos~\cite{Buckley:2010vg}. The diffuse gamma ray flux measurement
from portions of the sky~\cite{Cirelli:2009dv,Papucci:2009gd} as well as
the isotropic near full-sky
\cite{Abazajian:2010sq,Abdo:2010dk,*Hutsi:2010ai,*Arina:2010rb} have
also been used to set bounds on dark matter annihilation.

Observational limits on the strengths of gamma ray lines can be
translated into bounds on the overall rate of dark matter annihilation
into monochromatic gamma rays. Such constraints have been obtained
using the data from Fermi-LAT~\cite{Abdo:2010nc,Abdo:2010nz},
\cite{Vertongen:2011mu}, as well as earlier data from
EGRET~\cite{Pullen:2006sy} and other observations~\cite{Mack:2008wu}.
However, in contrast to the case of continuum gamma rays, these limits
have primarily been used to probe specific well-motivated models, rather
than to obtain general bounds on the overall rate of dark matter
annihilation into various SM final states.

The reason for this is that in typical models of WIMP dark matter, the
relic abundance is set by tree level annihilation to final states
consisting of two SM fermions or two weak gauge bosons.  In this
scenario monochromatic gamma rays corresponding to annihilation to the
$\gamma \gamma$, $\gamma Z$ and $\gamma H$ final states are only
generated at one loop. The strengths of the corresponding gamma ray
lines are highly model dependent, being very sensitive to the detailed
structure of the corresponding matrix element, which in turn depends on
the details of the vertices in the theory. Furthermore, the amplitude
will in general receive significant contributions from unknown non-SM
states (associated with new physics) running in the loop. This makes it
challenging to translate the observational bounds on gamma ray lines
into model-independent constraints on the rate of dark matter
annihilation or decay. Instead, limits have been set on specific models
motivated either by the hierarchy problem or by simplicity. In
particular, WIMP models for which the one loop cross sections to the
$\gamma \gamma$ and $\gamma Z$ final states have been computed include
the Minimal Supersymmetric Standard Model
(MSSM)~\cite{Bergstrom:1997fh,Bern:1997ng,Ullio:1997ke,Boudjema:2005hb}
and its extension to include a
singlet~\cite{Ferrer:2006hy,Chalons:2011ia}, Universal Extra Dimensions
(UED)~\cite{Bergstrom:2004nr,Bertone:2009cb,Bertone:2010fn}, the Littlest
Higgs Model with T-Parity~\cite{Birkedal:2006fz,Perelstein:2006bq}, the
Inert Doublet Model~\cite{Gustafsson:2007pc} and the Scalar Singlet Dark
Matter Model~\cite{Profumo:2010kp}. In theories where dark matter is a
scalar or Majorana fermion, annihilation to the $\gamma H$ final state
is suppressed in the non-relativistic limit. However, the cross section
to this and to the $\gamma Z$ final state have been computed in a
specific theory of Dirac dark matter~\cite{Jackson:2009kg} that arises
in a class of Randall-Sundrum models. A model-independent approach based
on effective operators that correlates the signal in direct detection
experiments with the approximate line strength has also been
developed~\cite{Goodman:2010qn}, (see also~\cite{Badin:2009cf}).

In this paper, we place lower bounds on the strengths of gamma ray lines
from dark matter annihilation that are applicable to a large class of
theories. We focus on models where the photon spectrum includes both
continuum gamma rays that arise from annihilation into two body SM final
states at tree level, as well as monochromatic gamma rays arising from
annihilation directly into two photons at the one loop level. For this
class of theories we obtain lower limits on the ratio of the strength of
the gamma ray line relative to the gamma ray continuum as a function of
the dark matter mass and spin. These limits are obtained from the
unitarity relation between the tree level amplitude of the primary
annihilation channel and the imaginary part of the loop level amplitude
for annihilation directly into photons, with the primary decay products
running in the loop.  These bounds are exact in the limit that dark
matter annihilates exclusively into a single SM species, that the
annihilation is dominated by the $L=0$ partial wave and respects $CP$.
Relaxing these assumptions renders our limits approximate, but does not
invalidate them. In particular, we stress that these results do not
depend on the specific form of the coupling between the dark matter
particle and the SM field running in the loop, and are in that sense
model-independent.  Our conclusions agree with known results in the
literature for the MSSM, UED and the Littlest Higgs Model with T-Parity.
We use the Fermi-LAT observations to translate these limits into upper
bounds on the dark matter annihilation cross section into any specific
SM state.

\section{Lower limits on Line strengths}

In this section, we explain in detail how these limits are obtained.
Unitarity implies that the $S$-matrix satisfies
\begin{align}
S^{\dagger}S = 1.
\end{align}
Writing $S = 1 + iT$, where $T$ is the transition matrix, we obtain
\begin{align}
-i(T - T^{\dagger}) = T^{\dagger} T.
\end{align}
Consider the matrix element of this equation between the initial state
$|i \rangle $ consisting of the two dark matter particles, and the final
state $|f \rangle $ consisting of two photons,
\begin{align}
- i \langle f | (T - T^{\dagger}) | i \rangle =
\displaystyle\sum\limits_X \langle f | T^{\dagger} | X \rangle
\langle X | T | i \rangle,
\label{unitarity1}
\end{align}
where the sum over $ | X \rangle $ runs over all final states into
which the dark matter particle can annihilate.

In the center of mass frame, two particle states can be labelled by the
internal quantum numbers of the particles, the total energy, and the
angular momentum quantum numbers $ | J, M; L, S \rangle $, where $J$ is
the total angular momentum, $M$ is the component of total angular
momentum along any fixed axis, $L$ is the orbital angular momentum and
$S$ the total spin angular momentum. Dark matter annihilation in haloes
occurs in the highly non-relativistic regime. We therefore focus on the
case where annihilation occurs through the $L = 0$ partial wave, so that
the initial state $| i \rangle $ has $J =S$. While $J$ and $M$ are
conserved in the annihilation process, this is not true of $L$ or $S$.
If $CP$ is a symmetry of the theory, states composed of a particle and
its antiparticle, or of two self-conjugate particles, are eigenstates of
$CP$ in this basis. The corresponding eigenvalue is $(-1)^S$ if the
particles are bosons and $ - (-1)^S$ if they are fermions.

There is an alternative basis for labelling two particle states in the
center of mass frame, the helicity basis, which will also prove useful.
In this basis, in addition to the internal quantum numbers of the
particles and their total energy, the states are labelled by the angular
momentum quantum numbers $ | J, M; \lambda_1, \lambda_2 \rangle $, where
$\lambda_1$ and $\lambda_2$ are the helicities of the two particles.
Unlike $J$ and $M$, the helicities are not conserved in the annihilation
process. The transformation that relates the $ | J, M; \lambda_1,
\lambda_2 \rangle $ basis to the $ | J, M; L, S \rangle $ basis may be
found in the classic paper by Jacob and Wick \cite{Jacob:1959at}.

Now, if the theory is invariant under time reversal (or equivalently
$CP$), and the states $|i \rangle $ and $|f \rangle $ are eigenstates of
angular momentum, labelled either by $ | J, M; L, S \rangle $ or by $ |
J, M; \lambda_1, \lambda_2 \rangle $, then, as shown in the appendix,
$\langle f | T | i \rangle = \langle i | T | f \rangle $.
Eq.~(\ref{unitarity1}) then simplifies to
 \begin{align}
2  {\rm Im} \langle f | T | i \rangle =
\displaystyle\sum\limits_X \langle f | T^{\dagger} | X \rangle
\langle X | T | i \rangle.
 \end{align}
At lowest order in perturbation theory, this equation relates the
imaginary part of the loop amplitude for annihilation into the two
photon final state to the tree amplitude for annihilation into two body SM
states. Squaring, we get
\begin{align}
4 |{\rm Im} \langle f | T | i \rangle|^2  =
\left| \displaystyle\sum\limits_X
\langle f | T^{\dagger} | X \rangle \langle X | T | i \rangle \right|^2.
\end{align}

We now restrict ourselves to the case where dark matter annihilation at
tree level is exclusively to a single SM species. If the SM state $|X
\rangle $ is further characterized by fixed values of $L$ and $S$ (or
fixed values of $\lambda_1$ and $\lambda_2$) for any given values $J$
and $M$ of the initial state $| i \rangle$, then
 \begin{align}
4 |{\rm Im} \langle f | T | i \rangle|^2  =
|\langle f | T^{\dagger} | X \rangle |^2 |\langle X | T | i \rangle|^2.
\end{align}
\begin{figure*}[htp]
  \begin{center}
    \begin{align*}
    \frac{
    \sigma_{IM}\left(
    \includegraphics[scale=0.7,trim=0.0cm 1.2cm 0cm 0cm ]{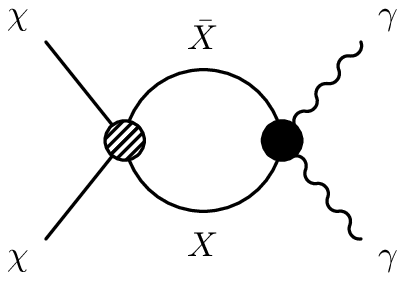}
    \right)
    }{
    \hspace{3mm}
    \sigma\left(
    \includegraphics[scale=0.7,trim=0.0cm 0.8cm 0cm 0cm ]{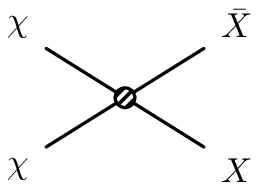}
    \right)
    }
    \quad&=\quad
    \frac{
    \Gamma_{\rm Im}\left(
    \includegraphics[scale=0.7,trim=0.0cm 1.2cm 0cm 0cm ]{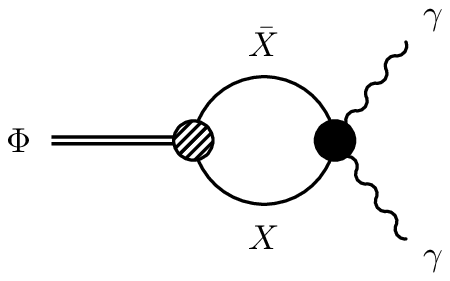}
    \right)
    }{
    \hspace{3mm}
    \Gamma\left(
    \includegraphics[scale=0.7,trim=0.0cm 0.8cm 0cm 0cm ]{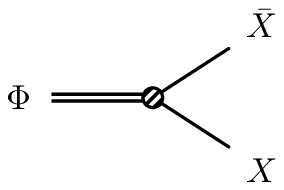}
    \right)
    }
    \end{align*}
  \end{center}
  \caption{
  Model independence of the lower bound on the ratio of loop and
  tree annihilation cross sections shown graphically.
  The $X$ particle is a SM
  state, and $\chi$ is the dark matter particle. For purposes of
  obtaining a limit, dark matter annihilation can be modeled by
  the decay of a boson, $\Phi$.
  \label{fig:ann-decay}
  }
\end{figure*}
This equation implies that for all initial states $ | i \rangle $ that
annihilate at tree level exclusively to the SM state $| X \rangle $
characterized by fixed $L$ and $S$ (or fixed $\lambda_1$ and
$\lambda_2$), the ratio
 \begin{align}
\frac{|{\rm Im} \langle f | T | i \rangle|^2}{|\langle X | T | i \rangle|^2}
= \frac{1}{4} |\langle f | T | X \rangle|^2
\end{align}
is a constant that depends only on SM parameters, and is otherwise
independent of $| i \rangle$. Furthermore, under these conditions the
quantity
\begin{align}
\frac{|\langle f | T | i \rangle|^2}{|\langle X | T | i \rangle|^2}
\geq \frac{1}{4} |\langle f | T | X \rangle|^2
\end{align}
is bounded from below. The numerator of the expression on the left hand
side is proportional to the cross section for $ |i \rangle \rightarrow
|f \rangle$, while the denominator is proportional to the cross
section for $ | i \rangle \rightarrow | X \rangle$. The right hand side
is proportional to the absolute value of the tree level matrix element
for annihilation of the state $| X \rangle $, consisting of two SM
particles, into two photons. This can be calculated within the SM.
Therefore this expression can be translated into a lower bound on the
rate for $ \chi \chi \rightarrow \gamma \gamma $ relative to the rate
for $ \chi \chi \rightarrow X \; \bar{X} $. To obtain the bound we
proceed as follows.

\begin{itemize}
\item
We construct a theory with a bosonic particle $\Phi$ of mass $m_{\Phi} = 2
m_{\chi}$ that decays exclusively into the state $ | X \rangle$ at tree level,
and into two photons at one loop. The spin and $CP$ properties of $\Phi$ are
chosen to be identical to those of the initial state composed of two dark
matter particles.
\item
We calculate the contribution to the decay rate for $\Phi \rightarrow
\gamma \gamma$ that arises purely from the imaginary part of the
corresponding one loop amplitude. We denote this quantity by
$\Gamma_{\rm Im} (\Phi \rightarrow \gamma \gamma)$.
\item
We calculate the decay rate for $\Phi \rightarrow X \; \bar{X}$, denoted by
$\Gamma (\Phi \rightarrow X \; \bar{X})$.
\item
As shown above, the ratio of these two decay rates will be
independent of the details of interactions of $\Phi$.  Therefore, the
decays of $\Phi$ can be used to model dark matter annihilation. Hence,
we obtain the bound (see Fig. \ref{fig:ann-decay}):
\begin{align}
\frac{\sigma( \chi \chi \rightarrow \gamma \gamma)}
{\sigma( \chi \chi \rightarrow X \; \bar{X})} \geq
\frac{\Gamma_{\rm Im} (\Phi \rightarrow \gamma \gamma)}
{\Gamma (\Phi \rightarrow X \; \bar{X})}.
\end{align}
\item
Using the fact that the continuum spectrum of gamma rays corresponding to
dark matter annihilation to the state $|X \rangle$ is known, this
result can be translated into a lower bound on the strength of the
gamma ray line relative to the continuum.
\end{itemize}

For the bound to apply, dark matter annihilation at tree level must
occur exclusively into a single SM final state. Furthermore, this state
must be characterized by definite values of $L$ and $S$ (or definite
$\lambda_1$ and $\lambda_2$), for any given $J$ and $M$. Under what
circumstances are these criteria satisfied? This turns out to depend on
the dark matter spin, on whether the dark matter particle is its own
anti-particle, and on the masses ands spins of the SM particles in the
state $|X \rangle$. We now consider the various possibilities in turn.
\\

\noindent
{\underline{Scalar dark matter}}

Consider first the case where dark matter is a scalar, either real or
complex. Then the initial state is $CP$ even, and has $J =0$. Angular
momentum conservation implies that annihilation into light fermions is
helicity suppressed. Therefore we expect that annihilation will occur
primarily to the heaviest SM fermion species that is kinematically
accessible, or alternatively to $W^+ W^-$ (excluding neutral states
which obviously do not contribute to photon line signal at one loop).
In the case of annihilation to fermions, angular momentum and $CP$
conservation imply that the final state must have $L =1, S =1$, while
$L = 0, S =0$ is forbidden.  Therefore a bound can be obtained for
this annihilation mode.

For annihilation to $W^+ W^-$ on the other hand, the conservation laws
allow both $L = 0, S =0$ and $L =2, S =2$ final states, forbidding only $L
=1, S =1$. Therefore our formalism is not directly applicable.
Nevertheless, in the next section we shall see that in various kinematic
limits a bound can indeed be obtained.
\\

\noindent
{\underline{Majorana fermion dark matter}}

We move on to the case where dark matter is a Majorana fermion. Since
Majorana fermions are identical particles, anti-symmetry of their wave
function implies that if $L =0$, $S$ is also then zero so that $J =0$.
However, unlike the case of scalar dark matter, the initial state is now $CP$
odd. As before annihilation to light fermions is disfavored by angular
momentum considerations, and so the heavy fermion and $W^+ W^-$ final
states are again expected to dominate. Consider first annihilation to
fermions. Angular momentum and $CP$ conservation imply that the final state
must have $L =0, S =0$, while $L =1, S= 1$ is forbidden. Therefore our
formalism applies to this annihilation channel. What about the $W^+W^-$
channel? Now the conservation laws allow only the $L=1, S =1$ final state,
while forbidding the $L =0, S= 0$ and $L =2, S =2$ final states. Therefore
our formalism applies to this channel as well.
\\

\noindent
{\underline{Dirac fermion dark matter}}

If dark matter is a Dirac fermion there are two possibilities for the
total angular momentum of the initial state, $J =0$ or $J =1$.
Annihilation will in general proceed through both these channels, and
since $J$ is a conserved quantum number there is no interference. The
bound in each channel can be calculated independently. The weaker of
these two limits is then the true bound, corresponding to the case where
annihilation occurs entirely through that channel.  For $J = 0$, $CP$ is
odd and the analysis is identical to that of Majorana fermion dark
matter considered above. However, if $J =1$, annihilation to the $\gamma
\gamma$ final state is forbidden by the Landau-Yang
theorem~\cite{Landau,Yang:1950rg}. Therefore
dark matter can annihilate to SM final states through this channel
without giving rise to a line signal from the two photon final state. We
conclude that in the case that dark matter is a Dirac fermion, no
general bound is possible. \\

\noindent
{\underline{Vector boson dark matter}}

Finally we consider the case where dark matter is a real vector boson.
Since the initial state is composed of identical particles, the wave
function must be symmetric, allowing $J =0$ or $J =2$, but forbidding $J
=1$. Both $J = 0$ and $J = 2$ are $CP$ even. For $J =0$ the analysis is
identical to that of scalar dark matter considered above. However, for
$J = 2$ a separate analysis is needed. Annihilation may occur to either
light or heavy fermions, or to $W^+ W^-$. In particular, annihilation to
light fermions is no longer disfavored by angular momentum conservation.
Annihilation to fermions may occur either through any of the
$L=\{1,2,3\}, S=1$  channels. Therefore, even if annihilation
occurs exclusively to a single fermion species, it is not possible to
obtain a general bound. However, it is possible to obtain a bound in
various kinematic limits.

Annihilation to $W^+ W^-$ can also occur through multiple channels,
including $L=2, S=0$, and $L=\{0,1,2,3,4\}, S=2$.
Therefore, in general it is not possible to constrain
this channel either. However, we shall see that it is possible to obtain a
bound in the limit that the dark matter mass is close to the $W$ mass, so
that the final state $W$ bosons are non-relativistic.

\begin{table*}[htp]
  \begin{center}
    %\begin{tabular}{c|c|c|c|c|c}
      \begin{tabular}{
        |>{\centering}m{1.2in}|
         >{\centering}m{0.65in}|
         m{0.5in}<{\centering}|
         m{1.4in}<{\centering}|
         m{1.2in}<{\centering}|}
      \hline
      \multirow{2}{*}{Dark Matter}&
      \multirow{2}{*}{Initial spin}&\multicolumn{2}{c|}{Annihilation}
      &\multirow{2}{*}{Bound}\\
      \cline{3-4}
      & & Channel&Mode&\\
      \hline
      \rule{0pt}{10pt}
      \multirow{3}{*}{Scalar}
      &\multirow{3}{*}{$J=0$}
      & \multirow{2}{*} {$WW$} &$L=0, S=0$ &
      \multirow{2}{1.3in}
      {\centering In
      NR / UR limits.}
      %{\centering In
      %non-relativistic, ultra-relativistic limits.}
      \\
      &                     &&$L=2, S=2$ &   \\
      \cline{3-5}
      \rule{0pt}{15pt}
      && $f\bar{f}$  & $L=1, S=1$ & \checkmark \\
      \hline
      \rule{0pt}{15pt}
      \multirow{2}{*}{Majorana Fermion}
      &\multirow{2}{*}{$J=0$}
      & $WW$ &$L=1, S=1$ & \checkmark \\
      \cline{3-5}
      \rule{0pt}{15pt}
      && $f\bar{f}$  & $L=0, S=0$ & \checkmark \\
      \hline
      \rule{0pt}{15pt}
      \multirow{3}{*}{Dirac Fermion}
      &\multirow{2}{*}{$J=0$}
      & $WW$ &$L=1, S=1$ &\checkmark\\
      \cline{3-5}
      \rule{0pt}{15pt}
      && $f\bar{f}$  & $L=0, S=0$ &\checkmark \\
      \cline{2-5}
      \rule{0pt}{15pt}
      &$J=1$&\multicolumn{3}{c|}{Forbidden}\\
      \hline
      \rule{0pt}{10pt}
      \multirow{7}{*}{Real Vector Boson}
      &\multirow{3}{*}{$J=0$}
      & \multirow{2}{*} {$WW$} &$L=0, S=0$ &
      \multirow{2}{1.3in}
      {\centering In
      NR / UR limits.}
      \\
      &                     &&$L=2, S=2$ &\\
      \cline{3-5}
      \rule{0pt}{15pt}
      && $f\bar{f}$  & $L=0, S=0$ &\checkmark \\
      \cline{2-5}
      \rule{0pt}{10 pt}
      &\multirow{4}{*}{$J=2$}
      & \multirow{2}{*} {$WW$} &$L=2, S=0$ &\multirow{2}{*}{In
      NR limit.}\\
      &                     &&$L=\{0,1,2,3,4\}, S=2$ &   \\
      \cline{3-5}
      \rule{0pt}{15pt}
      && {$f\bar{f}$}  & $L=\{1,2,3\}, S=1$
      &
      {\centering In
      NR / UR limits.}
      \\
      \hline
    \end{tabular}
  \end{center}
  \caption{Summary of the bounds corresponding to the different dark
  matter candidates and annihilation channels. We have also indicated
  cases where a bound is only applicable when the tree-level
  annihilation products (on-shell intermediate states in the loop) are
  non-relativistic (NR) or ultra-relativistic (UR).}
  \label{tab:setup}
\end{table*}

\section{Computation of limits}

In this section we place lower limits on line strengths by explicitly
calculating the ratio of the decay rates of the boson $\Phi$:
 \begin{equation}
\frac{\Gamma_{\rm Im} (\Phi \rightarrow \gamma \gamma)}
{\Gamma (\Phi \rightarrow X \; \bar{X})}.
 \end{equation}
As outlined earlier, different choices of the spin and $CP$ properties
of the boson $\Phi$ map on to different dark matter candidates. We
choose the mass of $\Phi$ to be $2 m_{\chi}$, to ensure that the
intermediate state particles have exactly the same total energy as in
dark matter annihilation. This is required so that the quantum numbers
of the intermediate state can be exactly the same in the two cases,
which is required for the mapping to go through.
The final result will in general be seen to depend on the
velocity $\beta$ of the intermediate state particle $X$,
\begin{align}
  \beta
  &=
  \sqrt{1-\frac {m_{X}^2}{m_{\chi}^2}}.
\end{align}

\subsection{Scalar Dark Matter}

In the case where the dark matter particle is a scalar, the incoming
state is restricted to be in a $J=0$ state for non-relativistic
annihilation. The initial state is then a $CP$ even state with zero total
angular momentum, which is duplicated when the boson $\Phi$ is a scalar
particle, which we label by $\phi$.
\\

\noindent
{\underline{Scalar dark matter annihilation to fermions}}

As explained in the introduction, symmetry considerations require the
fermions to be in the $L=1, S=1$ final state. We assume that dark matter
annihilates predominantly to a single species of SM fermions $f \bar{f}$.
In the absence of large new sources of chiral symmetry breaking in the
theory, this is a safe assumption, since the matrix element for
annihilation a given fermion species is proportional to the fermion mass.
Then annihilation to heavier fermions such as tops and bottoms are preferred,
provided those channels are kinematically open.

We begin from the following interaction Lagrangian, which is assumed
to be valid below
the scale of electroweak symmetry breaking:
\begin{align}
  \mathcal{L}_{int}
  &=
  \lambda\, \bar{f} f\, \phi.
 \end{align}
From this Lagrangian it is straightforward to calculate the ratio of the
imaginary part of the decay to two photons to the total decay rate to
$f\bar{f}$. We obtain for the bound
\begin{align}
  \frac{\Gamma_{\rm Im}(\phi\to\gamma\gamma)
  }{
  \Gamma(\phi\to f\bar{f})
  }
  &=
  \frac{ N_c Q^4 e^4 m_f^2 }{32\pi^2 m_{\chi}^2}
  %\sqrt{1-\frac{4m_f^2}{s}}
  \beta
  \left[
  \tanh^{-1} \beta
  %\log\left[ \frac{x_+}{x_-}\right]
  \right]^2,
\label{scalarstofermionsbound}
\end{align}
where $Q$ is the electric charge of the fermion (e.g. $Q=\frac23$ for
top quarks) and $N_c = 3$ is the color factor to be included when the
fermionic states are quarks.
\\

\noindent
{\underline{Scalar dark matter annihilation to $W$ bosons}}

As explained earlier, in this case annihilation can proceed through
either the $L =0, S =0$ channel or through the $L =2, S =2$ channel.
Therefore it is only possible to obtain model independent bounds in
specific kinematic limits.

Consider first the limit that the dark matter mass is close to the $W$
mass, $m_\chi - m_W \ll m_W$. Then the annihilation products will be
non-relativistic and we expect that the $L = 0, S =0$ final state will
dominate. Therefore, in this limit a bound can be obtained. As before we
model the annihilation by the decay of a scalar particle $\phi$. We choose
the coupling of $\phi$ to $W$ bosons to have the simple form
\begin{align}
  \mathcal{L}_{int} &=
  \frac{1}{\Lambda} \phi \; \text{Tr}
\left[ F_{\mu\nu} F^{\mu\nu} \right],
\end{align}
and calculate the decay rates to $WW$ at tree level, and to $\gamma \gamma$
at loop level. The required ratio of decay rates is given by
\begin{align}
  \frac{\Gamma_{\rm Im}(\phi\to\gamma\gamma)
  }{
  \Gamma(\phi\to WW)
  }
  &=
  \frac{3 e^4}{64\pi^2 }
  \beta
  \label{eq:phiWW}
  \\
  \nonumber
  &\quad\text{(non-relativistic limit)}.
\end{align}
This is the bound in the non-relativistic limit.

In the opposite limit where the dark matter mass is much larger than the
$W$ mass, $m_\chi \gg m_W$, the $W$ bosons are ultra-relativistic.
Then the Goldstone boson equivalence theorem applies, and the
longitudinal and transverse components of the $W$ bosons correspond to
distinct physical states. Annihilation can occur either to two
longitudinal $W$ bosons or to two transverse $W$ bosons, and it is
reasonable to assume that there is no large cancellation in the
contributions to the $\gamma \gamma$ amplitude from these different
states. We therefore consider them separately. (Annihilation to one
transverse and one longitudinal $W$ is forbidden by angular momentum
conservation, and so need not be considered.)

If annihilation is exclusively to longitudinal $W$'s, then there is a
unique final state labelled by $ |0, 0, 0, 0 \rangle$ in the helicity
basis $| J, M, \lambda_1, \lambda_2 \rangle$, and our formalism can
immediately be applied. If annihilation is exclusively to transverse
$W$'s, then angular momentum conservation allows both the $ | 0, 0, +, +
\rangle $ and $ | 0, 0, -, - \rangle $ final states. However, since the
action of $CP$ reverses the helicities of single particle states, only
one linear combination of these states is $CP$ even, while the other is
$CP$ odd. Since annihilation can only occur to the $CP$ even state, once
again our formalism is applicable.

We first consider annihilation to longitudinal $W$ bosons. We couple the
scalar $\phi$ to the Higgs doublet, and consider decays into charged
Higgses, and into two photons through a loop of charged Higgses. The
Goldstone boson equivalence theorem guarantees that in the
ultra-relativistic limit, the rates for these processes are identical to
the rates for the corresponding processes involving $W$ bosons. The
interaction Lagrangian takes the form
\begin{align}
  \mathcal{L}_{int} &=
  \alpha \phi \; H^{\dagger} H.
\end{align}
The ratio of the decay rates in this case turns out to be suppressed.
It scales roughly as the following,
\begin{align}
  \frac{\Gamma_{\rm Im}(\phi\to\gamma\gamma)
  }{
  \Gamma(\phi\to WW)
  }
  &\sim
  \frac{e^4}{16\pi^2}\frac{m_W^4}{m_{\chi}^4}
  \left[
  \log \left(
  \frac{ 4 m_{\chi}^2}{m_W^2}
  \right)
  \right]^2.
  \\
  &\nonumber
  \qquad\qquad
  \text{(ultra-relativistic limit)}.
\end{align}
Since this vanishes in the limit $m_W^2/m_{\chi}^2 \rightarrow 0$,
this shows that there is in fact no bound in the case of annihilation to
longitudinal $W$ bosons in the ultra-relativistic limit.

We now consider annihilation to the transverse polarizations. We start
from the interaction Lagrangian
\begin{align}
  \mathcal{L}_{int} &=
  \frac{1}{\Lambda} \phi \; \text{Tr}
\left[ F_{\mu\nu} F^{\mu\nu} \right].
\end{align}
In the ultra-relativistic limit, this interaction leads to decays of the
scalar $\phi$ exclusively into transverse polarizations of the $W$ boson.
The ratio of cross-sections is obtained as
\begin{align}
  \frac{\Gamma_{\rm Im}(\phi\to\gamma\gamma)
  }{
  \Gamma(\phi\to WW)
  }
  &=
  \frac{e^4}{32\pi^2 }
  \left[
  \log \left(
  \frac{ 4 m_{\chi}^2}{m_W^2}
  \right)
  \right]^2
  \\
  &\nonumber\qquad\qquad\text{(ultra-relativistic limit)}.
\end{align}
This is then the lower bound in the case of annihilation exclusively into
transverse polarizations of the $W$.

In general, annihilation will occur into both transverse and longitudinal
polarizations. From the discussion above, it follows that if F$_{\rm T}$ is
the branching fraction into transverse polarizations of the $W$ in the
ultra-relativistic limit, the bound is given by
\begin{align}
\label{scalarstorWbound}
  \frac{\Gamma_{\rm Im}(\phi\to\gamma\gamma)
  }{
  \Gamma(\phi\to WW)
  }
  &= {\rm F}_{\rm T}
  \frac{e^4}{32\pi^2 }
  \left[
  \log \left(
  \frac{ 4 m_{\chi}^2}{m_W^2}
  \right)
  \right]^2
  \\
  &\nonumber\qquad\qquad\text{(ultra-relativistic limit)}.
\end{align}
This formula is valid in the ultra-relativistic limit provided
 \begin{equation}
{\rm F}_{\rm T} \gg \frac{{m_W}^2}{m_{\chi}^2}.
 \end{equation}
If the branching fraction to transverse polarizations is smaller than this,
then there is no bound in the limit $m_W^2/m_{\chi}^2 \rightarrow 0$.

\subsection{Majorana Fermion Dark Matter}

In the case when the dark matter particles are Majorana fermions, the
initial state again has total angular momentum $J=0$. This is because in
the non-relativistic limit $L=0$ is picked out, and then $S=0$ is
required by the overall antisymmetry of the total wavefunction of the two
fermion state. This configuration is $CP$ odd. The angular momentum and
$CP$ quantum numbers of the initial state are exactly those of a
pseudo-scalar particle. Therefore we can calculate the ratio by decaying a
pseudo-scalar particle, which we denote by $\varphi$.
\\

\begin{figure}[t]
\includegraphics[width=3.4truein]{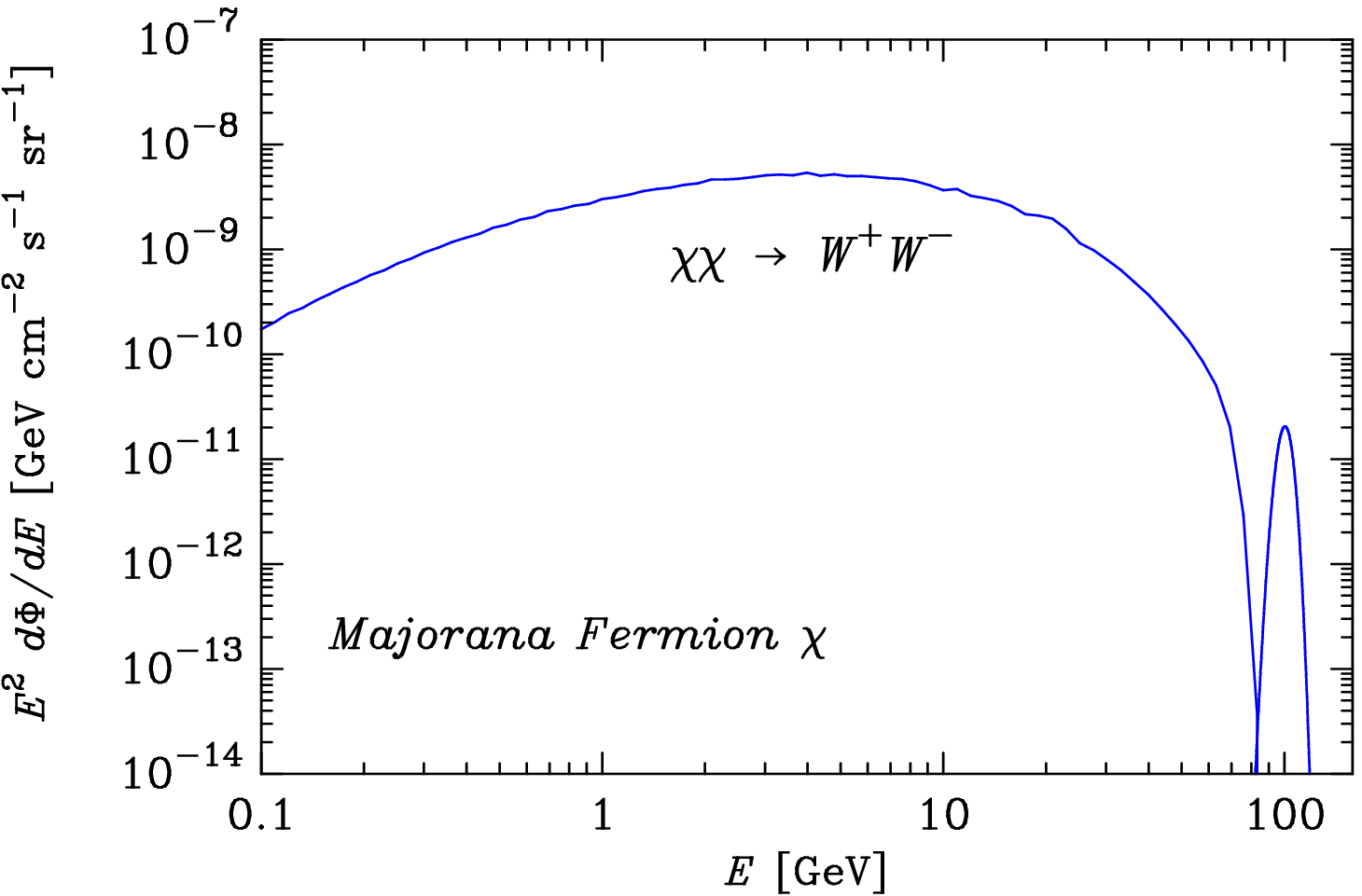}\\\ \\
\includegraphics[width=3.4truein]{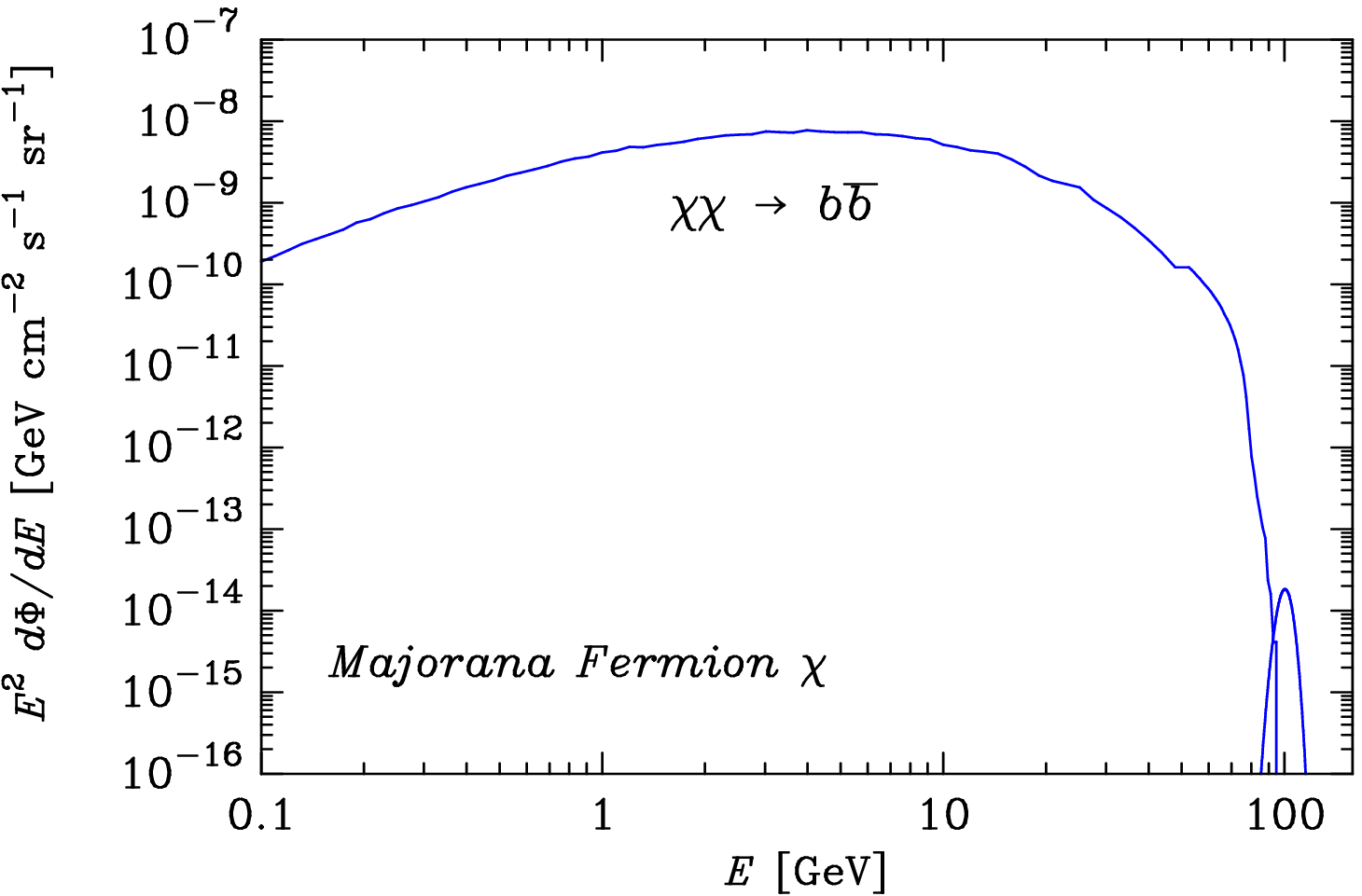}
\caption{Spectral flux $E^2 d\Phi/dE$ in the continuum relative to the
lower limit for the line obtained from our calculation.  We show
results for annihilation into $WW$ (upper panel) and $b\bar{b}$~(lower
panel) for a Majorana fermion dark matter. The annihilation
cross section was taken to be $\langle\sigma_{\rm A}v\rangle =
3\times10^{-26}\rm\ cm^3\,s^{-1}$ and $m_\chi = 100\rm\ GeV$.  The
width of the line is set by the Fermi-LAT resolution at 100 GeV, 11\%
FWHM.
\label{lineplot}
}
\end{figure}
\noindent
{\underline{Majorana dark matter annihilation to fermions}}

We begin by considering annihilation to fermions. Once again we assume
annihilation is exclusively to a single SM fermion species. As in the case
of scalar dark matter this is a safe assumption, since annihilation to
light fermions is chirality suppressed in the absence of large new sources
of chiral symmetry breaking in the theory.

Angular momentum conservation and $CP$ symmetry require the final state
fermions to be in the $L=0, S=0$ configuration rather than $L =1, S =1$.
Our starting point is the following interaction Lagrangian which couples the
pseudo-scalar $\varphi$ to the fermions $f \bar{f}$:
\begin{align}
  \mathcal{L}_{int}
  &=
  i\lambda\bar{f}\, \gamma^5 f\, \varphi.
\end{align}
From explicit calculation we obtain the bound in this case as
\begin{align}
  \frac{\Gamma_{\rm Im}(\varphi\to\gamma\gamma)
  }{
  \Gamma(\varphi\to f\bar{f})
  }
  &=
  \frac{N_c Q^4 e^4 m_f^2 }{32\pi^2 m_{\chi}^2}
  %\left[1 - \frac{4m_f^2}{s}\right]^{-1/2}
  \frac{1}{\beta}
  \left[
  \tanh^{-1}\beta
  %\log\left[ \frac{x_+}{x_-}\right]
  \right]^2,
\label{mfermionstofermionsbound}
\end{align}
where as before $Q$ denotes the electric charge, and $N_c$ denotes the
color factor associated with the fermion. We have verified that this
agrees with the result in the literature for the case of neutralino
annihilation to fermions in the
MSSM~\cite{Bergstrom:1997fh,Bern:1997ng}.
\\

\noindent
{\underline{Majorana dark matter annihilation to $W$ bosons}}

We move on to considering annihilation to $W$ bosons. The only angular
momentum quantum numbers for the $WW$ final state consistent with $J =0$
and $CP$ conservation are $L=1, S=1$. As before, we model this annihilation
by pseudo-scalar decay. The pseudo-scalar $\varphi$ is coupled to the SM in
the following way:
\begin{align}
  \mathcal{L}_{int} &=
  \frac{1}{\Lambda} \varphi
  \text{Tr}(F_{\mu\nu} \tilde{F}^{\mu\nu}).
\end{align}
This leads to the bound:
\begin{align}
  \frac{\Gamma_{\rm Im}(\varphi\to\gamma\gamma)
  }{
  \Gamma(\varphi\to WW)
  }
  &=
  \frac{e^4}{8\pi^2 }
  \beta
  \left[
  \tanh^{-1}\beta
  \right]^2.
\end{align}
This agrees with the result in the literature for neutralino annihilation
to $W$ bosons in the MSSM~\cite{Bergstrom:1997fh,Bern:1997ng}.

Fig.~\ref{lineplot} shows the lower limits on the strengths of the
gamma ray line relative to the gamma ray continuum if Majorana
fermion dark matter annihilates exclusively to the $b \bar{b}$ and
$WW$ final states respectively. The widths of the emission lines are
drawn so as to correspond with the resolution of the Fermi-LAT, at
11\% of FWHM.

To obtain the continuum DM annihilation spectra plotted in
fig.~\ref{lineplot}, we use Pythia 6.4~\cite{Sjostrand:2006za} to
simulate both photon radiation off of charged particles as well as
secondary photons from decays of particles such as the $\pi^{0}$.
Specifically, we run Pythia to simulate collisions at a center of mass
energy of exactly $2m_{\chi}$ through a Higgs boson decaying to a
final state that corresponds to the annihilation products. We switch
off initial state radiation to ensure that all photons are emitted
either radiatively off the final state particles or arise from the
decays of unstable particles such as mesons.  We use the default
Pythia cutoff values for photon emission and turn on the decays of
particles which are not decayed with the default Pythia settings, such
as muons, charged pions and kaons.

\subsection{Real vector boson dark matter}

Real vector boson dark matter annihilation can proceed through either the
$J=0$ channel or the $J=2$ channel. The $J=1$ channel is not allowed
because the wavefunction is required to be symmetric under interchange of
the two identical dark matter particles. Both the $J=0$ and $J=2$ initial
states are $CP$ even.

If dark matter annihilation occurs primarily through the $J=0$ channel,
then the bounds from scalar dark matter annihilation to the
corresponding SM states, Eqs.~{\ref{scalarstofermionsbound}},
{\ref{eq:phiWW}} and {\ref{scalarstorWbound}}, are relevant here as
well. We have verified that Eqs.~{\ref{scalarstofermionsbound}} and
{\ref{eq:phiWW}} agree with the
literature~\cite{Birkedal:2006fz,Perelstein:2006bq} in the case of the
vector dark matter candidate in the Littlest Higgs Model with T-Parity.
(It turns out that Eq.~{\ref{scalarstorWbound}} does not apply to this
model because for heavier dark matter masses, annihilation is primarily
to longitudinal W bosons.) On the other hand, if the primary mode of
annihilation is through the $J=2$ channel, then obtaining the bound
involves calculating the decays of a spin-2 $CP$ even particle. The
model independent lower limit on the strength of the line corresponds to
the weaker of the bounds obtained in the $J=0$ and $J=2$ cases.

We use the decays of a massive spin-2 graviton to model the $J=2$
annihilation process. The massive graviton is taken to couple to the
stress energy tensor of the matter fields, in analogy with the couplings
of the Kaluza-Klein graviton in extra-dimensional theories
\cite{Giudice:1998ck,Han:1998sg}.
\\

\noindent
{\underline{Vector dark matter annihilation to fermions}}

We begin by considering annihilation to fermions. The bounds we obtain
assume annihilation exclusively to a single species of SM fermion. Away
from this limit our bounds are approximate. The two fermion final states
that can arise from annihilation of the $J = 2$ initial state, after taking
into account angular momentum and $CP$ conservation, have angular momentum
quantum numbers $L = 1, S =1$, $L =2, S =1$ and $L =3, S =1$. This
multiplicity of available states means that it is only possible to obtain a
bound in specific kinematic limits.

We first consider the limit where the dark matter mass is close to the mass
of the fermion species it annihilates into, so that the outgoing fermions
are non-relativistic. Then the $L=1, S =1$ final state dominates, and
our formalism is applicable.

Consider a massive Dirac fermion $f$ that couples to a massive graviton
$h_{\mu \nu}$. The term in the Lagrangian which is relevant for the
on-shell graviton decay is the following:
\begin{align}
  \mathcal{L}_{int}
  &=
  -\frac{\kappa}{2}
  h^{\mu\nu}
  \bar{f} \;i\gamma_\mu \partial_\nu f.
\end{align}
The coupling constant $\kappa$ has dimensions of inverse mass. The mass of
the fermion couples to the trace of the graviton $h^\mu_\mu$, which
vanishes for the on-shell decay. The limit in this case is obtained very
weak because of the p-wave suppression of the amplitude:
\begin{align}
  \left.
  \frac{\Gamma_{\rm Im}(h \to\gamma\gamma)
  }{
  \Gamma(h \to f\bar{f})
  }\right|_{J=2}
  &=
  \frac{N_c Q^4 e^4\beta^3}{120\pi^2 }
  \label{eq:J2tt}
  \\&\nonumber\qquad\text{(non-relativistic limit)}.
\end{align}
This is significantly weaker than the corresponding bound on decays from
the $J=0$ state.  We conclude that for real vector dark matter annihilation
to heavy fermions in the non-relativistic limit, it is the bound from the
$J = 2$ initial state that applies.

Next we consider the case of annihilation to very light SM fermions,
such as electrons or muons. In particular, we work in the limit where
the dark matter mass is much larger than the mass of the final state
fermions. Then the final state fermions are ultra-relativistic, and can
be treated as massless. In this limit, left- and right-handed fermions
of the same flavor are effectively different SM species, and should be
considered separately. We now show that in the ultra-relativistic limit,
for annihilation exclusively to a single SM fermion species of definite
chirality, it is possible to obtain a bound. To understand how this
arises we work in the helicity basis, and consider annihilation
exclusively to left-handed Weyl fermions (and right-handed
anti-fermions). The unique final state is then $| J, M; +\frac12,
-\frac12
\rangle $, and so our formalism can immediately be applied.

The couplings of a massless chiral Weyl fermion $f$ to a massive
graviton take the form:
\begin{align}
  \mathcal{L}_{int}
  &=
  -\frac{\kappa}{2}
  h^{\mu\nu}
  \bar{f} \;i \bar{\sigma}_\mu \partial_\nu f.
\end{align}
We obtain the result
\begin{align}
  \left.
  \frac{\Gamma_{\rm Im}(h \to\gamma\gamma)
  }{
  \Gamma(h \to f\bar{f})
  }\right|_{J=2}
  &=
  \frac{N_c Q^4 e^4}{144\pi^2 }
  \label{eq:J2ff}
  \\&\nonumber\qquad\text{(ultra-relativistic limit)}.
\end{align}
Since annihilation to massless fermions from the initial state with $J =
0$ is forbidden, this is the applicable bound. This agrees with the
result in the literature~\cite{Bergstrom:2004nr,Bertone:2010fn} for the
case of the vector dark matter candidate in UED annihilating to light
chiral fermions.

In this scenario, in contrast to the cases of scalar and Majorana
fermion dark matter, the annihilation cross section to SM fermions is
not chirality suppressed. It is then very natural for dark matter to
couple identically to all three flavors of a SM (chiral) fermion species
and annihilate with equal strength to all of them. In the absence of other
annihilation modes, Eq.~\ref{eq:J2ff} gets modified to
\begin{align}
  \left.
  \frac{\Gamma_{\rm Im}(h \to\gamma\gamma)
  }{
  \Gamma(h \to f\bar{f})
  }\right|_{J=2}
  &=
  \frac{N_f N_c Q^4 e^4}{144\pi^2 }
  \\&\nonumber\qquad\text{(ultra-relativistic limit)}.
\end{align}
where $N_f$ is the number of flavors.
\\

\noindent
{\underline{Vector dark matter annihilation to $W$ bosons}}

For annihilation to $W$ bosons from the $J =0$ initial state, the earlier
bounds from the case of scalar dark matter apply in the appropriate
kinematic limits. In the case of the $J =2$ initial state, even after
imposing angular momentum and $CP$ conservation, there are still several
channels that can contribute. It is possible to obtain a bound in the limit
that the dark matter mass is close to the $W$ boson mass, $m_{\chi} - m_W
\ll m_W$, when annihilation to the $L =0, S =2$ final state dominates.

The relevant part of the Lagrangian, in unitary gauge, takes the form:
\begin{align}
  \mathcal{L}_{int}
  &= \frac{\kappa}{2} h^{\mu\nu}
  \left(
  \left[
  (\partial_\mu W^{+\rho} - \partial^\rho W^+_\mu)
  (\partial_\nu W^-_\rho - \partial_\rho W^-_\nu)
  \right.\right.\nonumber\\&\left.\left.
  \qquad\qquad\qquad
  - m_W^2 W^+_\mu W^-_\nu
  \right]
  + \mu \leftrightarrow \nu
  \right)
\end{align}
Explicit calculation leads to the bound:
\begin{align}
  \left.
  \frac{\Gamma_{\rm Im}(h\to\gamma\gamma)
  }{
  \Gamma(h\to WW)
  }\right|_{J=2}
  &=
  \frac{e^4}{20\pi^2 }
  \beta
  \label{eq:J2WW}
  \\&\nonumber\qquad\text{(non-relativistic limit)}.
\end{align}
This bound is (slightly) stronger than than that obtained for
annihilation through the $J=0$ channel (Eq. \ref{eq:phiWW}). We
conclude that the $J=0$ channel provides the appropriate bound
for vector dark matter annihilation to $WW$ in the non-relativistic limit.

\section{Significance of the Results}

\begin{figure}[t]
\includegraphics[width=3.2truein]{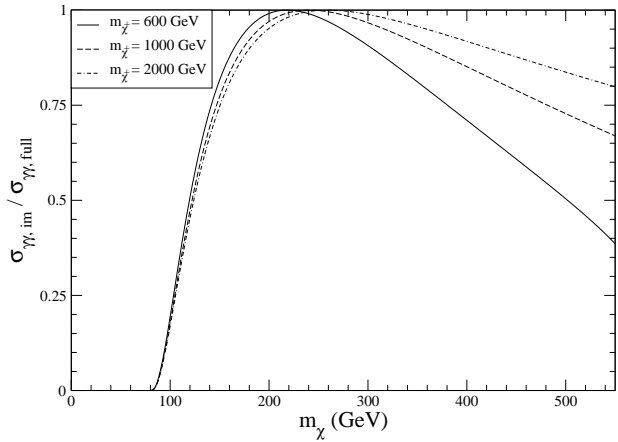}
\caption{For the SUSY model of ~\cite{Bergstrom:1997fh,Bern:1997ng}
where the neutralino DM annihilates through a loop of $W$-bosons into a
pair of photons, we plot the ratio of the part of the annihilation cross
section arising from the imaginary part of the amplitude to the full
cross section.}
\label{ratioplotSUSY}
\end{figure}

In order for these results to be useful, and to lead to meaningful
bounds on the dark matter annihilation cross section, the contribution
of the imaginary part of the amplitude to the line strength must be
comparable to the contribution of the real part of the amplitude. While
in general we expect that this will be the case, in certain
circumstances the imaginary part of the amplitude may be suppressed
relative to the real part as a consequence of symmetry considerations.
One important example of such a suppression occurs when dark matter
annihilates from a $J = 0$ initial state into two photons through a loop
of light fermions. If the intermediate state fermions $f$ and $\bar{f}$
are on shell, their subsequent annihilation into two photons is
chirality suppressed. (This is in addition to the chirality suppression
of the initial annihilation process into two fermions itself.) As a
consequence, the contribution of the imaginary part of the amplitude to
the dark matter cross section to two photons is suppressed by the square
of the ratio of the fermion mass $m_f$ to the dark matter mass
$m_{\chi}$.
 \begin{equation}
\frac{\sigma_{\rm Im}( \chi \chi \rightarrow \gamma \gamma)}
{\sigma( \chi \chi \rightarrow f \bar{f})}
\propto
\left[\frac{m_f}{m_{\chi}}\right]^2
 \end{equation}
This is evidenced in Eqs.~{\ref{scalarstofermionsbound}} and
{\ref{mfermionstofermionsbound}}. There is in general no corresponding
suppression for the real part of the amplitude. Therefore we expect that
the limits we obtain in these cases will be significantly weaker than those
obtained from the full amplitude.

\begin{figure}[t]
\includegraphics[width=3.2truein]{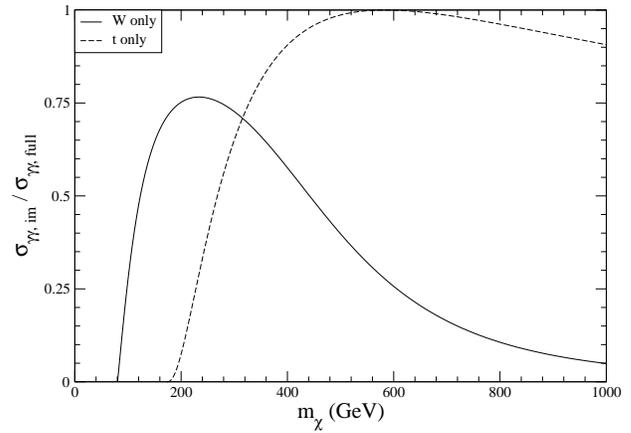}
\caption{For the Littlest Higgs Model with T-parity~\cite{Birkedal:2006fz,Perelstein:2006bq}
where the DM annihilates into a pair of photons through an s-channel Higgs
boson, we plot the ratio of the part of the annihilation cross section
arising from the imaginary part of the amplitude to the full cross section.
The contribution purely due to the W's in the loop and purely due to
the top quark in the loop is shown.}
\label{ratioplotLH}
\end{figure}

\begin{figure}[t]
\includegraphics[width=3.2truein]{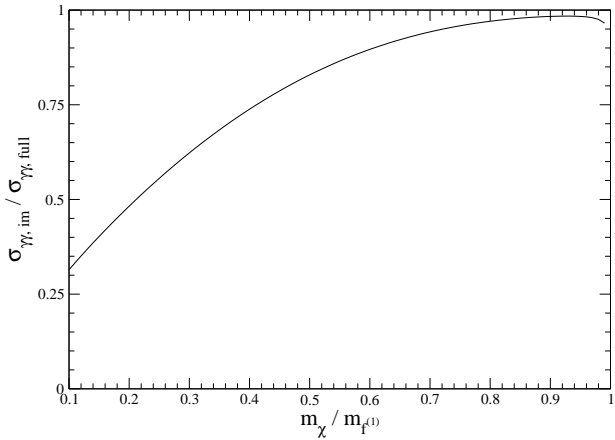}
\caption{For the UED model of~\cite{Bergstrom:2004nr,Bertone:2009cb} where the
KK-photon DM annihilates into a pair of photons through a loop of light
fermions, we plot the ratio of the part of the annihilation cross section
arising from the imaginary part of the amplitude to the full cross
section as a function of the ratio of the KK-photon mass to the
KK-fermion mass.}
\label{ratioplotUED}
\end{figure}

In order to obtain an understanding of the relative sizes of the
contributions of the imaginary part of the amplitude and the real part
of the amplitude to the strength of the gamma ray line in realistic
theories, we compute the ratio
 \begin{equation}
\mathcal{R} =
\frac{\sigma_{\rm Im}( \chi \chi \rightarrow \gamma \gamma)}
{\sigma( \chi \chi \rightarrow \gamma \gamma)}
 \end{equation}
as a function of the dark matter mass for a few specific well-motivated
models, in the limit that tree level annihilation of $\chi$ is to a
specific SM final state.  The results are plotted in Figs.
\ref{ratioplotSUSY}, \ref{ratioplotLH} and \ref{ratioplotUED}.

In Fig.~\ref{ratioplotSUSY} we consider neutralino annihilation into two
photons through a loop of $W$ bosons in the MSSM. The initial
annihilation into two $W$ bosons is mediated by the chargino. We have
therefore plotted $\mathcal{R}$ as a function of the neutralino mass for
several different values of the chargino mass. The ratio $\mathcal{R}$
for this case is easily obtained from
\cite{Bergstrom:1997fh,Bern:1997ng}. It is clear from the figure that
the contribution of the imaginary part of the amplitude is significant.

In Fig.~\ref{ratioplotLH} we consider dark matter annihilation in the
Littlest Higgs Model with T-parity
\cite{Birkedal:2006fz,Perelstein:2006bq}. We first study annihilation
through a loop of top quarks and then through a loop of $W$-bosons. Of
course, in the complete model, the contributions of the top quarks and
the $W$'s cannot be individually turned on or off as the annihilation
proceeds through the Higgs boson, which has sizable couplings to both
states. Nevertheless, our plot illustrates that the imaginary part of
the amplitude contributes significantly to the total cross section.

In Fig.~\ref{ratioplotUED} we consider Kaluza-Klein photon dark matter
annihilation to a pair of photons through a loop of chiral fermions in
UED. The initial annihilation process of $\chi$ into two SM fermions is
mediated by Kaluza-Klein fermions. We have computed $\mathcal{R}$ as a
function of the dark matter mass, for a given Kaluza-Klein fermion mass,
using the results from~\cite{Bergstrom:2004nr,Bertone:2010fn}. Once
again we see that the imaginary part of the amplitude contributes a
sizable fraction of the total annihilation cross section.

Our results show that in general the contribution of the imaginary part
of the amplitude is expected to be very significant. In fact, in the
cases we studied, it dominates the cross section in a large part of the
interesting parameter space. We conclude from this that, except in those
cases where symmetry considerations suppress the imaginary part of the
amplitude, the bounds we obtain will in general not be significantly
weaker than those obtained from the full amplitude.

\section{Current and Future Experimental Limits}
\label{sec:limits}

In this section we combine these results with the current experimental
constraints on gamma ray lines obtained by the Fermi-LAT
collaboration~\cite{Abdo:2010nc} to place upper bounds on the rate of
dark matter annihilation into various SM final states, and estimate
potential future sensitivities.

We consider each of scalar, Majorana fermion and real vector boson dark
matter in turn and consider annihilation into a few plausible SM final
states. We obtain a lower limit on the strength of the photon line given
the annihilation cross section into each final state. The null result in
Fermi-LAT line search then allows us to put an upper bound on the
corresponding annihilation cross section. Our results are then compared
to the corresponding bounds on dark matter annihilation obtained from
the Fermi-LAT constraints on continuum gamma rays. This allows us to
determine the relative strength of the bounds from the line and the
continuum photon spectra searches.

In obtaining these limits, we have not taken into account the
distortions to the spectrum in the neighbourhood of the line that
would be caused by the continuum gamma rays that necessarily arise
from the same dark matter annihilation channel. This can be a very
significant effect, particularly in the case of annihilation into
light SM fermions~\cite{Bergstrom:2004cy},\cite{Birkedal:2005ep},
\cite{Bringmann:2011ye}, (see also \cite{Bringmann:2012vr}). To
understand the size of this effect, in Table~\ref{tab:photon-bins} we
have shown the probability of obtaining a photon of a given energy
from annihilation of a pair of Majorana fermion dark matter particles
of mass 100 GeV.  We have considered annihilation into $b \bar{b}$ and
$W^{+}W^{-}$ final states. The table shows the resulting spectrum of
continuum gamma rays in each case, and a line with strength
corresponding to the lower bound.

The data has been separated into bins of
energy 5 GeV, with and without taking into account the effects of the
energy resolution of Fermi-LAT, modeled here with a Gaussian of FWHM
10 GeV. It is clear from the table that in the case of the $b \bar{b}$
final state, the line and the continuum are not very well separated
once the resolution is taken into account, so that picking out the
line may be challenging. This is reflected in the fact that the
fraction of events associated with the line in the bins centered at 90
GeV, 95 GeV and 100 GeV is $13.7\%$, $77.0\%$ and $97.3\%$
respectively.
However, in the case of the $W^{+}W^{-}$
final state, the line and continuum are well separated, and the
effects of the continuum spectrum in the neighbourhood of the line are
quite small even allowing for the finite energy resolution. 
The fraction of events associated with the line in the bins centered at 90
GeV, 95 GeV and 100 GeV is now a much more healthy $94.7\%$, $99.9\%$
and $100.0\%$
respectively.

More generally, the cases where the effect of the continuum spectrum
in the neighborhood of the line is important correspond to scenarios
where annihilation is primarily to light fermions. However, as will be
clear from the limits, in these cases the bound arising from the
continuum gamma rays is currently much stronger than the one from the
line. If the energy resolution of the experiments were to improve to
the point where the limit arising from lines became comparable to that
from the continuum, the analysis for these cases would have to be
redone incorporating this effect in order to obtain a completely
realistic bound. We leave this for future work.

\begin{table*}[tp] \vspace{0.3in}
  \begin{center}
    \renewcommand{\tabcolsep}{12pt}
    \subfloat[][]{
    \begin{tabular}{lll}
      \hline \hline
      \multirow{2}{*}{Bin (GeV)}& \multicolumn{2}{c}{Probability per
      event} \\
      & Perfect Resolution	 & Finite Resolution\\
      \hline
      $60 $  &  $3.21\times10^{-4}$ & $3.24\times10^{-4}$\\
      $65 $  &  $1.71\times10^{-4}$ & $1.77\times10^{-4}$\\
      $70 $  &  $8.03\times10^{-5}$ & $8.60\times10^{-5}$\\
      $75 $  &  $2.77\times10^{-5}$ & $3.40\times10^{-5}$\\
      $80 $  &  $3.49\times10^{-6}$ & $9.56\times10^{-6}$\\
      $85 $  &  $3.92\times10^{-7}$ & $1.74\times10^{-6}$\\
      $90 $  &  $6.80\times10^{-8}$ & $2.87\times10^{-7}$\\
      $95 $  &  $2.00\times10^{-9}$ & $1.63\times10^{-7}$\\
      $100$  &  $5.27\times10^{-7}$ & $1.90\times10^{-7}$\\
      $105$  &  $0.00             $ & $1.26\times10^{-7}$\\
      $110$  &  $0.00             $ & $3.95\times10^{-8}$\\
      $115$  &  $0.00             $ & $5.69\times10^{-9}$\\
      \hline
      \hline
    \end{tabular}
    }
    %\qquad\qquad
    \subfloat[][]{
    \begin{tabular}{lll}
      \hline \hline
      \multirow{2}{*}{Bin (GeV)}& \multicolumn{2}{c}{Probability per
      event} \\
      & Perfect Resolution	 & Finite Resolution\\
      \hline
      $60$   &$3.13\times10^{-4}   $  & $3.22\times10^{-4}$ \\
      $65$   &$1.78\times10^{-4}   $  & $1.80\times10^{-4}$ \\
      $70$   &$7.95\times10^{-5}   $  & $8.66\times10^{-5}$ \\
      $75$   &$2.74\times10^{-5}   $  & $3.40\times10^{-5}$ \\
      $80$   &$3.00\times10^{-6}   $  & $9.53\times10^{-6}$ \\
      $85$   &$1.00\times10^{-7}   $  & $1.94\times10^{-6}$ \\
      $90$   &$0.00                $  & $2.78\times10^{-6}$ \\
      $95$   &$0.00                $  & $8.40\times10^{-6}$ \\
      $100$  &$3.52\times10^{-5}   $  & $1.23\times10^{-5}$ \\
      $105$  &$0.00                $  & $8.39\times10^{-6}$ \\
      $110$  &$0.00                $  & $2.63\times10^{-6}$ \\
      $115$  &$0.00                $  & $3.80\times10^{-7}$ \\
      \hline
      \hline
    \end{tabular}
    }
  \end{center}
  \caption{Photon yield probabilities (per annihilation event) for
  Majorana dark matter of mass 100 GeV annihilating into a) $b\bar{b}$
  and b) $WW$ in different energy bins with and without the effects of the Fermi-LAT energy resolution.}
  \label{tab:photon-bins}
\end{table*}

The flux in a line is given by
\begin{align}
  \frac{d\Phi}{dE}
  &=
  \frac{\langle \sigma_{\rm A} v\rangle}
  {8\pi m_\chi^2}
  \frac{\mathcal{J}}{{\rm J}_0}
  \frac{dN}{dE},
\end{align}
where
\begin{align}
  \frac{dN}{dE} = 2 \delta(E_\gamma - m_\chi)
\end{align}
 for the given annihilation
channel with annihilation cross section $\langle \sigma_{\rm A}
v\rangle$, and ${\rm J_0} \equiv 1/[8.5\ \rm kpc (0.3\ GeV\
cm^{-3})^2]$.  The normalized integral of the mass density squared
over the observational regions for the line search is
 \begin{equation}
  \mathcal{J} =
{\rm J_0} \int{\rho^2\left(r_{\rm gal}(b,\ell,x)\right)\cos b\ dx\ db\ d\ell},
  \label{jcal}
 \end{equation}
where $r_{\rm gal}(b,\ell,x)$ is the radial coordinate of the density
distribution.  The regions included in the line search are the Galactic
caps at $|b| > 10^\circ$ and the $20^\circ\times 20^\circ$ region in the
Galactic center at $|b| < 10^\circ$ and $|\ell| < 10^\circ$.  To arrive
at our line annihilation cross section limits, we use the line flux
limit from the second column of Table I of Ref.~\cite{Abdo:2010nc} to
place limits that employ a consistent and conservative choice for the
Milky Way dark matter halo for both the line search and diffuse flux
limits.  We have verified that we arrive at the same annihilation
cross-sections in the remaining entries of Table I of
Ref.~\cite{Abdo:2010nc} when we adopt the designated dark matter halo
(e.g., NFW). In our derived limits here, we adopt a minimal model of the
Milky Way Galactic dark matter profile which minimizes the signal yet is
consistent with constraints from Galactic dynamics and is consistent
with profiles expected in cold dark matter halo formation.  To do so, we
take an Einasto profile for the dark matter density distribution with
parameters that minimize $\mathcal{J}$ and are consistent with the
dynamical constraints in Ref.~\cite{Catena:2009mf}.  The Einasto profile
is
 \begin{align}
  \rho_{\rm Einasto} (r) &=
  \rho_s
  \exp\left[-\frac{2}{\alpha}
  \left(
  \left[\frac{r}{r_s}\right]^\alpha -1
  \right)
  \right],
 \end{align}
and the parameters from Ref.~\cite{Catena:2009mf} within 68\%
confidence level (CL) that give a lower $\mathcal J$ are $\alpha =
0.22$, $r_s = 21\rm\ kpc$, with $r_\odot = 8.28\rm\ kpc$ and $r_S$
determined from the local solar dark matter density $\rho_\odot =
0.385\rm\ GeV\ cm^{-3}$.  The value of the normalized density squared
toward the Galactic caps in this case is $\mathcal{J}_{\rm cap} =
28.6$ and toward the Galactic Center's $20^\circ$ square is
$\mathcal{J}_{\rm GC} = 37.0$.

\begin{figure*}[ht]
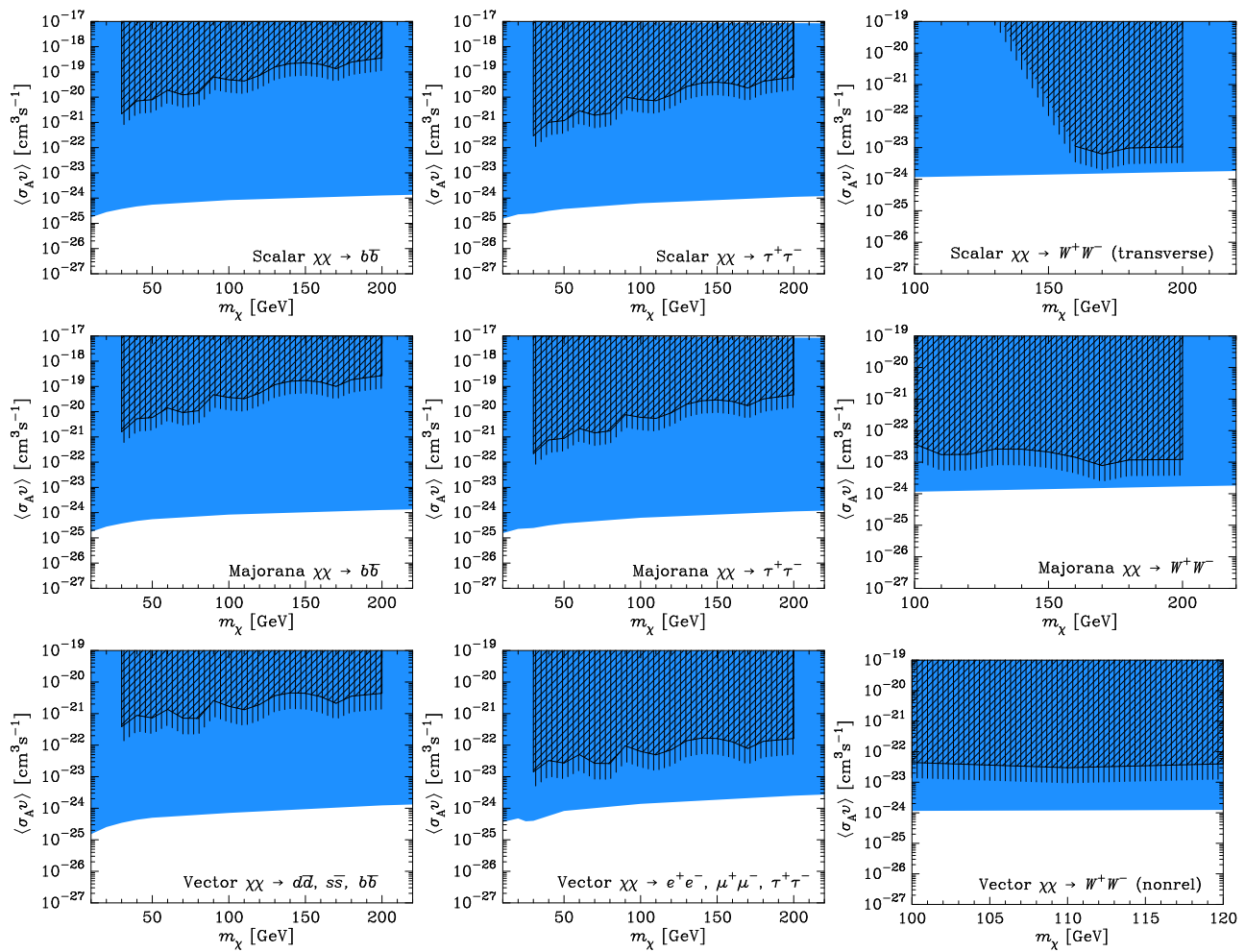

\mbox{\includegraphics[width=2.2truein]{bbbar_line_scalar.eps}
\includegraphics[width=2.2truein]{tautau_line_scalar.eps}
\includegraphics[width=2.2truein]{ww_line_scalartrans.eps}}\\
\mbox{\includegraphics[width=2.2truein]{bbbar_line_majorana.eps}
\includegraphics[width=2.2truein]{tautau_line_majorana.eps}
\includegraphics[width=2.2truein]{ww_line_majorana.eps}}\\
\mbox{\includegraphics[width=2.2truein]{qqbar_line_vector.eps}
\includegraphics[width=2.2truein]{3l_line_vector.eps}
\includegraphics[width=2.2truein]{ww_line_vector.eps}}
\caption{\small Upper, middle and lower rows of panels show current
  and future 95\% CL constraints on dark matter annihilation channels
  in the scalar, Majorana, and vector dark matter cases respectively.
  The colored regions are from constraints from the isotropic DGRB on
  Galactic and extragalactic dark matter annihilation, as described in
  the text.  The diagonally hatched regions are from current
  constraints from Fermi-LAT on the minimal line contribution in the
  respective channels, as discussed in the text. The vertically
  hatched regions are forecasts for the limits after 10 years of
  observation with Fermi-LAT, assuming Poisson statistics at the
  current energy resolution. In the case of scalar dark matter to $WW$
  (upper right panel), the solid line indicates the true limit for
  ultra-relativistic $WW$ production, and the non-outlined hatched
  region is an interpolation of the limit to the non-relativistic
  regime at $m_\chi\sim 100\rm\ GeV$.
  \label{limitfig}}
\end{figure*}

The calculated line limits depend on the adopted profile among
dynamically consistent models at the level of $\sim$50\%, as shown in
Ref.~\cite{Abdo:2010nc}. Our adopted profile and parameters give a
minimal $\mathcal{J}$, and therefore provide a conservative limit that
is consistent with the profiles used to derive the diffuse flux
continuum spectrum limits, and our choice of halo profile is consistent
with cold dark matter halo structure formation. If we were to choose the
exact NFW or Einasto profiles as that in Ref.~\cite{Abdo:2010nc}, it
would alter our present and forecast line constraints by less than
$50\%$.  Because we would like to show relative constraints within a
given conservative choice of a halo profile, we adopt the profile of the
minimal $\mathcal{J}_{\rm GC}$ for both the line search regions and the
DGRB continuum constraints.

In the case of Majorana fermion dark matter we consider annihilation
to the $b \bar{b}$, $ \tau^+ \tau^-$ and $WW$ final states. For scalar
dark matter we consider exactly the same final states, but with the
added restriction in the $WW$ case that the $W$-bosons are
ultra-relativistic and transverse. In the case of real vector boson
dark matter, we first consider the case where annihilation occurs
equally to all three generations of ultra-relativistic right-handed
down-type quarks $b \bar{b}$, $s \bar{s}$ and $d \bar{d}$, then the
case where it occurs equally to all three generations of
ultra-relativistic right-handed charged leptons $\tau \bar{\tau}$,
$\mu \bar{\mu}$ and $e \bar{e}$ and finally the case where
annihilation is to non-relativistic $W$-bosons.  The 95\% confidence
level (CL) limits are shown in Fig.~\ref{limitfig}.

The continuum limits, as shown in Fig.~\ref{limitfig}, are derived as
in Ref.~\cite{Abazajian:2010sq} from Galactic and extragalactic
contributions to the isotropic diffuse gamma ray background (DGRB) as
measured by Fermi-LAT~\cite{Abdo:2010nz}, with the exception that here
we take the substructure boost to be a very conservative $B =2.3$
within the Galactic halo component of the DGRB contribution.  This
boost value is found from the substructure enhancement as in
Ref.~\cite{Abazajian:2010sq}, with minimal substructure parameters,
and with a partial cancellation of the total luminosity boost
contribution from our position within the Galactic halo.

The minimal line limits are in all cases weaker than continuum limits
at this time from Fermi-LAT's measurement of the isotropic DGRB.  Note
that limits from the stacking of dwarf galaxies are stronger than that
from the DGRB by a factor of $\sim$10
\cite{GeringerSameth:2011iw,*Ackermann:2011wa} and the isotropic DGRB
sensitivity may be enhanced \cite{Abazajian:2010zb}.  We compare to
the isotropic DGRB constraints here because they are the most
conservative among the annihilation constraints.  In
 Fig.~\ref{limitfig} we also show the expected future 10-year Fermi-LAT
line limits assuming enhancement of the sensitivity with the exposure
time ($t$) of Fermi-LAT as $\sqrt{t}$.  The line sensitivity for Fermi-LAT may
improve substantially with systematic improvement at better than the
Poisson count rate. For example, this may occur with enhanced energy
resolution for lines over the lifetime of Fermi-LAT.  Moreover, for
certain cases of dark matter spins, any future gamma-ray experiment
with even higher energy resolution
(e.g. GAMMA-400~\cite{Galper:2011zz}) may prove to have line
sensitivities or constraints on dark matter annihilation be more
significant than that for the continuum.

\section{Conclusions}

In certain models where the dark matter particle has a primary tree
level annihilation channel to a unique final state in the SM, we have
derived a robust lower limit on the cross section of a loop-induced
annihilation mode to photons relative to the tree level annihilation
cross section to the specific SM state. This bound is based on
unitarity considerations that relate the imaginary part of the
loop-induced amplitude for annihilation to photons to the amplitude
for the primary annihilation channel. Since the spectrum of continuum
photons emitted through bremsstrahlung and decays of the primary
annihilation products can be reliably computed, this lower bound also
relates the minimal strength of a gamma ray line from dark matter
annihilation to the size of the continuum spectrum. While the very
conservative bounds for a gamma ray line obtained in this way are less
stringent than the bounds from the continuum photons, these results
help identify combinations of initial and final state quantum numbers
for which a model-dependent calculation of the full amplitude (rather
than only its imaginary part) for annihilation into photons can give
comparable bounds to those derived from continuum emission, and
constitute an independent consistency check on the results of any such
calculation.

\begin{acknowledgments}
It is a pleasure to thank Kaustubh Agashe, Raman Sundrum and Neal
Weiner for useful comments.  PA and ZC are supported by the NSF under
grant PHY-0801323 and PHY-0968854. KNA is supported by NSF Grant
07-57966 and NSF CAREER Grant 09-55415. CK is supported by the NSF
Grant Number PHY-0969020.
\end{acknowledgments}

\bigskip

\appendix \section{Time-Reversal Invariance and the $T$-Matrix}

In this appendix we show that if the theory respects time-reversal
invariance, then the $T$-matrix is symmetric in the angular
momentum basis,
 \begin{align}
\langle f | T | i \rangle = \langle i | T | f \rangle  \; ,
 \end{align}
where $| i \rangle$ and $| f \rangle$ are angular momentum eigenstates.

The fact that the theory is time-reversal invariant implies that
 \begin{equation}
\langle f | T | i \rangle = \langle \Theta i | T | \Theta f \rangle \;,
\label{TRdefinition}
 \end{equation}
where $\Theta$ is the time-reversal operator. The action of
time-reversal on eigenstates of angular momentum is given
by~\cite{Jacob:1959at}
 \begin{eqnarray}
{\Theta} |J,M ; L,S \rangle
  &=&
(-1)^{J-M} |J, -M ; L, S \rangle  \nonumber \\
{\Theta} |J,M ; \lambda_1, \lambda_2 \rangle
  &=&
(-1)^{J-M} |J, -M ; \lambda_1,  \lambda_2 \rangle  \; .
\label{TRonJM}
 \end{eqnarray}
Now rotational invariance of the theory, in the form of the
Wigner-Eckart theorem, implies that
 \begin{align}
  &\langle J'M';L'S' | T |J,M;LS\rangle
  \nonumber
  \\
  &
  \qquad\qquad\qquad =
  \delta_{J,J'} \delta_{M,M'}
  \langle L'S' | T^J |LS\rangle
  \nonumber\\&
  \qquad\qquad\qquad =
  \langle J',-M';L'S' | T |J,-M;LS\rangle \; .
\label{WignerEckart1}
 \end{align}
 \begin{align}
  &
  \langle J'M'; \lambda_1', \lambda_2'
  | T |
  J,M; \lambda_1, \lambda_2 \rangle
  \nonumber\\&
  \quad\qquad\qquad =
  \delta_{J,J'} \delta_{M,M'}
  \langle \lambda_1', \lambda_2' | T^J |\lambda_1, \lambda_2 \rangle
  \nonumber\\
  &
  \quad\qquad\qquad =
\langle J',-M';\lambda_1', \lambda_2' | T |J,-M;\lambda_1, \lambda_2 \rangle
\; .
\label{WignerEckart2}
 \end{align}
It follows from Eq.~(\ref{TRonJM}), Eq.~(\ref{WignerEckart1}) and
Eq.~(\ref{WignerEckart2}) that matrix elements between time-reversed
states obey
 \begin{align}
\langle \Theta i | T | \Theta f \rangle = \langle i | T | f \rangle \; ,
 \end{align}
where $| i \rangle$ and $| f \rangle $ are angular momentum eigenstates.
It then immediately follows from Eq~(\ref{TRdefinition}) that the
$T$-matrix is symmetric in this basis,
 \begin{align}
\langle f | T | i \rangle = \langle i | T | f \rangle \; .
 \end{align}

\bibliographystyle{apsrev4-1}
\bibliography{lines-ref}

\end{document}